\documentclass[twocolumn,floats,showpacs,superscriptaddress,pre]{revtex4}
\usepackage{amsmath,amssymb,bm}
\usepackage{graphics}
\usepackage{epsfig}
\usepackage{graphicx}

\begin{document}

\title{Quantum Geometry of a Configuration Space in a Covariant Dynamical
Theory}
\author{Natalia Gorobey, Alexander Lukyanenko}
\email{alex.lukyan@rambler.ru}
\affiliation{Department of Experimental Physics, St. Petersburg State Polytechnical
University, Polytekhnicheskaya 29, 195251, St. Petersburg, Russia}
\author{Inna Lukyanenko}
\email{inna.lukyanen@gmail.com}
\affiliation{Institut f\"{u}r Mathematik, TU Berlin, Strasse des 17 Juni 136, 10623
Berlin, Germany}

\begin{abstract}
A quantum version of the action principle in 
a simple covariant dynamical theory of two relativistic particles is
formulated. The central object of this new formulation of quantum theory is
a stationary eigenvalue of the quantum action. This quantity defines a
quantum geometry in a configuration space. 
In the presence of "probe" fields it plays the role of a generation function
of observables.
\end{abstract}

\maketitle
\date{\today }





\section{\textbf{INTRODUCTION}}

In the work \cite{GL} the formulation of quantum mechanics in terms of a
quantum version of the action principle was proposed. In this formulation
the action functional $I$ is replaced by an operator $\widehat{I}$\ in a
space of wave functionals. A wave functional $\Psi \left[ q\left( t\right) %
\right] $ is defined on trajectories $q\left( t\right) $ with fixed end
points in a configuration space of a dynamical system. The eigenvalue
problem of the action operator $\widehat{I}$ is formulated as follows:%
\begin{equation}
\widehat{I}\Psi =\Lambda \Psi .  \label{1}
\end{equation}%
Quantum dynamics of a system is described by an extremal eigenvalue $\Lambda
_{0}$ and a corresponding eigenfunctional $\Psi _{0}$ (the quantum action
principle). In the work \cite{GL1}, attention was paid to the fact that the
new formulation is the most appropriate for covariant systems. As an
example, the dynamics of a relativistic particle in the Minkowsky space was
considered. In contrast to the ordinary formulation of relativistic quantum
mechanics based on the Klein-Gordon equation, where the problem of
probabilistic interpretation of a wave function exists \cite{BD}, the
probabilistic interpretation of a wave functional $\Psi \left[ q\left(
t\right) \right] $ is natural. Incidentally, all symmetries of the original
classical theory - the relativistic invariance and the invariance with
respect to reparametrizations of a world line of a particle are preserved.

A special feature of the covariant theory is that the action, being
invariant with respect to reparametrizations of a trajectory, defines a
geometry which in general belongs to the class of the Finsler geometry \cite%
{F} in a configuration space of a system. In the case of a relativistic
particle it is the Minkowsky geometry. As the consequence, quantization of
the action introduces a quantum geometry of the configuration space. This
geometry is defined by the extremal eigenvalue $\Lambda _{0}$ of the action
which is a function of end points of a trajectory and, therefore, may be
taken as a measure of distance. Quantum geometry of the configuration space
defined in this way has direct physical meaning: it plays the role of a
generation function for mean values of currents and correlators of currents,
if the dynamics of system is "probed" by an external field. These quantities
form the set of physical observables.

In the present work a quantum geometry of a configuration space of a system,
that may be symbolically called the system of two particles, is considered.
According to the structure of the algebra of constraints this dynamical
system may serve as a simple model of General Relativity and string theories.

\section{COVARIANT SYSTEM OF TWO PARTICLES}

Classical dynamics of a system is described by a trajectory $\left( u^{\mu
}\left( \tau \right) ,v^{\mu }\left( \tau \right) \right) $ in the
configuration space $M_{4}\times M_{4}$, where $M_{4}$ is the
four-dimensional Minkowsky space, and $\tau \in \left[ 0,1\right] $ is an
arbitrary parameter. The dynamics is determined by the action:

\begin{eqnarray}
I &=&\int\limits_{0}^{1}\left[ \frac{1}{4N_{1}}\left( \overset{\cdot }{u}%
-\lambda u\right) ^{2}+\frac{1}{4N_{2}}\left( \overset{\cdot }{v}+\lambda
v\right) ^{2}\right.  \notag \\
&&\left. +N_{1}v^{2}+N_{2}u^{2}+A_{\mu }^{1}\left( u\right) \overset{\cdot }{%
u}^{\mu }+A_{\mu }^{2}\left( v\right) \overset{\cdot }{v}^{\mu }\right] d\tau
\label{2}
\end{eqnarray}%
where the dot denotes the derivative on the parameter $\tau $. Short
notations for scalar products and squares of vectors in the Minkowsky space
are used, for example, $u^{2}\equiv \eta _{\mu \nu }u^{\mu }u^{\nu }$, where

\begin{equation}
\eta _{\mu \nu }\equiv diag\left( +1,-1,-1,-1\right)  \label{3}
\end{equation}%
is the metric of the Minkowsky space. Apart from the basic dynamical
variables $\left( u^{\mu },v^{\mu }\right)$, the action (\ref{2}) depends on
$N_{1,2}$ and $\lambda $ which are analogous to lapse and shift functions in
the Arnowitt, Deser and Misner framework of General Relativity \cite{MTW}.
Transformation properties of these variables ensure the invariance of the
action with respect to reparametrizations of a trajectory. External "probe"
fields $A^{1}\left( u\right)$ and $A^{2}\left( v\right)$ are included for
the subsequent determination 
of observables (the charges of particles are equal to unity).

We obtain the classical geometry of the configuration space, induced by the
action (\ref{2}), by turning, first of all, to the geometrical form of the
action. For this purpose let us exclude lapse and shift functions from the
action (\ref{2}), solving corresponding Euler-Lagrange (EL) equations:

\begin{eqnarray}
-\frac{1}{4N_{1}^{2}}\left( \overset{\cdot }{u}-\lambda u\right) ^{2}+v^{2}
&=&0,  \notag \\
-\frac{1}{4N_{2}^{2}}\left( \overset{\cdot }{v}+\lambda v\right) ^{2}+u^{2}
&=&0,  \label{4} \\
-\frac{u\left( \overset{\cdot }{u}-\lambda u\right) }{4N_{1}}+\frac{v\left(
\overset{\cdot }{v}+\lambda v\right) }{4N_{2}} &=&0.  \notag
\end{eqnarray}%
After substitution of a solution of the set (\ref{4}) \ with respect to $%
N_{1,2},\lambda $\ into (\ref{2}), the action takes the geometrical form
with the integrand which is homogeneous function of the first order in the
velocities:

\begin{eqnarray}
I &=&\int\limits_{0}^{1}\left[ \sqrt{v^{2}}\sqrt{\left( \overset{\cdot }{u}%
-\lambda u\right) ^{2}}+\sqrt{u^{2}}\sqrt{\left( \overset{\cdot }{v}+\lambda
v\right) ^{2}}\right.  \notag \\
&&\left. +A^{1}\overset{\cdot }{u}+A^{2}\overset{\cdot }{v}\right] d\tau
\label{5}
\end{eqnarray}%
where

\begin{eqnarray}
\lambda &=&\frac{-b+\sqrt{b^{2}-4ac}}{2a},  \notag \\
a &\equiv &\left( v^{2}\right) ^{2}\left( \left( \overset{\cdot }{u}u\right)
^{2}-\overset{\cdot }{u}^{2}u^{2}\right) -\left( u^{2}\right) ^{2}\left(
\left( \overset{\cdot }{v}v\right) ^{2}-\overset{\cdot }{v}^{2}v^{2}\right) ,
\notag \\
b &\equiv &2\left[ \left( \overset{\cdot }{v}v\right) v^{2}\left( \left(
\overset{\cdot }{u}u\right) ^{2}-\overset{\cdot }{u}^{2}u^{2}\right) \right.
\notag \\
&&\left. +\left( \overset{\cdot }{u}u\right) u^{2}\left( \left( \overset{%
\cdot }{v}v\right) ^{2}-\overset{\cdot }{v}^{2}v^{2}\right) \right] ,  \notag
\\
c &\equiv &\left( \overset{\cdot }{u}u\right) ^{2}\overset{\cdot }{v}%
^{2}v^{2}-\left( \overset{\cdot }{v}v\right) ^{2}\overset{\cdot }{u}%
^{2}u^{2}.  \label{6}
\end{eqnarray}%
Considering the action (\ref{5}) as a measure of length of a trajectory $%
\left( u^{\mu }\left( \tau \right) ,v^{\mu }\left( \tau \right) \right)
,\tau \in \left[ 0,1\right] $, let us define a "shortness" trajectory from
the condition that the action (\ref{5}) is stationary with respect to the
basic variables, i.e., from the equations of motion of particles:

\begin{equation}
\delta _{u}I=\delta _{v}I=0.  \label{7}
\end{equation}%
The equations (\ref{7}) are differential equations of the second order in
the parameter $\tau $ with respect to functions $u^{\mu }\left( \tau \right)
,v^{\mu }\left( \tau \right) $. Substitution of a solution of these
equations at fixed end points of a trajectory, $\left( u_{0},v_{0}\right) $
and $\left( u_{1},v_{1}\right) $, into the action (\ref{5}) determines a
distance $I_{01}$\ between end points in a classical geometry of a
configuration space.

Taking into account that the quantity $I_{01}$ is a functional of "probe"
fields, let us define current density vectors of particles:

\begin{eqnarray}
\left. \frac{\delta I_{01}}{\delta A_{\mu }^{1}\left( u\right) }\right\vert
_{A=0} &\equiv &j_{1}^{\mu }\left( u\right) =\int\limits_{0}^{1}d\tau \delta
^{4}\left( u-u\left( \tau \right) \right) \overset{\cdot }{u}^{\mu }\left(
\tau \right) ,  \notag \\
\left. \frac{\delta I_{01}}{\delta A_{\mu }^{2}\left( v\right) }\right\vert
_{A=0} &\equiv &j_{2}^{\mu }\left( v\right) =\int\limits_{0}^{1}d\tau \delta
^{4}\left( v-v\left( \tau \right) \right) \overset{\cdot }{v}^{\mu }\left(
\tau \right) ,  \label{8}
\end{eqnarray}%
where integrals are taken on classical world lines of particles. Higher
order variational derivatives describe dependence of currents on external
fields. These quantities contain all details of internal interactions in a
system. In analogy with electrodynamics we call them correlators of
currents. These quantities form the complete set of observables of the
theory.

\section{CANONICAL FORM OF THE ACTION}

As usually, we begin the transition to a canonical form of the action,
defining canonical momenta of particles:

\begin{eqnarray}
p_{\mu } &\equiv &\frac{\partial L}{\partial u^{\mu }}=\frac{1}{2N_{1}}%
\left( \overset{\cdot }{u_{\mu }}-\lambda u_{\mu }\right) +A_{\mu }^{1},
\notag \\
\pi _{\mu } &\equiv &\frac{\partial L}{\partial v^{\mu }}=\frac{1}{2N_{2}}%
\left( \overset{\cdot }{v_{\mu }}+\lambda v_{\mu }\right) +A_{\mu }^{2}.
\label{9}
\end{eqnarray}%
Then we obtain the Hamiltonian:

\begin{equation}
H=\left. \left( p\overset{\cdot }{u}+\pi \overset{\cdot }{v}-L\right)
\right\vert _{\overset{\cdot }{u}\overset{\cdot }{v}}=N_{1}H_{1}+N_{2}H_{2}+%
\lambda D,  \label{10}
\end{equation}%
where velocities are supposed to be excluded by use of Eq. (\ref{9}), and

\begin{eqnarray}
H_{1} &\equiv &\left( p-A^{1}\right) ^{2}-v^{2}\approx 0,  \notag \\
H_{2} &\equiv &\left( \pi -A^{2}\right) ^{2}-u^{2}\approx 0,  \notag \\
D &\equiv &u\left( p-A^{1}\right) -v\left( \pi -A^{2}\right) \approx 0,
\label{11}
\end{eqnarray}%
are constraints of the first class. The wavy equalities mean that the
equations (\ref{11}), being the EL equations for $N_{1,2}$ and $\lambda $,
have to be solved, but only after calculation of all necessary Poisson
brackets. Among the latter are commutators of a Lee algebra of the
constraints:

\begin{equation}
\left\{ H_{1},H_{2}\right\} =4D,\left\{ D,H_{1}\right\} =2H_{1},\left\{
D,H_{2}\right\} =-2H_{2}.  \label{12}
\end{equation}%
In that case, the group of covariance is three-parametric. It includes
independent reparametrizations of world lines of particles.

Canonical quantum theory of a dynamical system in ordinary formulation was
considered in the work \cite{X}, where solutions of a set of wave equations:

\begin{equation}
\widehat{H}_{1}\psi =\widehat{H}_{2}\psi =\widehat{D}\psi =0,  \label{13}
\end{equation}%
for a wave function $\psi \left( u,v\right) $ were investigated. Operators
of the constraints in the Schr\"{o}dinger representation are obtained
by a replacement of the canonical momenta by differential operators acting
on a wave function:

\begin{equation}
\widehat{p}_{\mu }\equiv \frac{\hslash }{i}\frac{\partial }{\partial u^{\mu }%
},\widehat{\pi }_{\mu }\equiv \frac{\hslash }{i}\frac{\partial }{\partial
v^{\mu }}.  \label{14}
\end{equation}%
However, we are forced to note, once again, that any probabilistic
interpretation of solutions of the set (\ref{13}) is absent. Proper
probabilistic interpretation will be found in the new form of quantum theory
based on the quantum action principle.

\section{QUANTUM ACTION PRINCIPLE}

The central point \ in the modified procedure of the canonical quantization,
proposed in \cite{GL}, is the replacement of partial derivatives (\ref{14})
acting on a wave function $\psi \left( u,v\right) $ by variational
derivatives acting on a wave functional $\Psi \left[ u\left( \tau \right)
,v\left( \tau \right) \right] $. This functional also depends on the lapse
functions $N_{1,2}\left( \tau \right) $, but at present the latter are
assumed to be fixed. More precisely, the functional realization of basic
canonical variables is as follows:

\begin{eqnarray}
\widehat{u}^{\mu }\left( \tau \right) \Psi &\equiv &u^{\mu }\left( \tau
\right) \Psi ,\widehat{v}^{\mu }\left( \tau \right) \Psi \equiv v^{\mu
}\left( \tau \right) \Psi ,  \label{15} \\
\widehat{p}_{\mu }\left( \tau \right) \Psi &\equiv &\frac{\widetilde{\hslash
}}{i}\frac{\delta \Psi }{\delta u^{\mu }\left( \tau \right) },\widehat{\pi }%
_{\mu }\left( \tau \right) \Psi \equiv \frac{\widetilde{\hslash }}{i}\frac{%
\delta \Psi }{\delta v^{\mu }\left( \tau \right) }  \label{16}
\end{eqnarray}%
where the variational derivatives are defined as follows:

\begin{eqnarray}
\delta \Psi &=&\int\limits_{0}^{1}\left[ \frac{\delta \Psi }{\delta u^{\mu
}\left( \tau \right) }\delta u^{\mu }\left( \tau \right) N_{1}\left( \tau
\right) \right.  \notag \\
&&\left. +\frac{\delta \Psi }{\delta v^{\mu }\left( \tau \right) }\delta
v^{\mu }\left( \tau \right) N_{2}\left( \tau \right) \right] d\tau .
\label{17}
\end{eqnarray}%
The lapse functions $N_{1,2}$ are included into the measure of integration
to ensure the covariance of the new quantum theory. The constant $\widetilde{%
\hslash }$ differs from the "ordinary" Plank constant $\hslash $, its
physical dimensionality is $\left[ \widetilde{\hslash }\right] =Joule\cdot
s^{2}$. A relationship between two constants may be established after a
determination of observables and comparison between the theory and the
experiment. Operators defined in this way obey the following permutation
relations:

\begin{eqnarray}
\left[ \widehat{u}^{\mu }\left( \tau \right) ,\widehat{p}_{\nu }\left( \tau
^{^{\prime }}\right) \right] &=&i\widetilde{\hslash }\delta _{\nu }^{\mu }%
\frac{1}{N_{1}}\delta \left( \tau -\tau ^{^{\prime }}\right) ,  \notag \\
\left[ \widehat{v}^{\mu }\left( \tau \right) ,\widehat{\pi }_{\nu }\left(
\tau ^{^{\prime }}\right) \right] &=&i\widetilde{\hslash }\delta _{\nu
}^{\mu }\frac{1}{N_{2}}\delta \left( \tau -\tau ^{^{\prime }}\right) .
\label{18}
\end{eqnarray}%
They are formally Hermitian with respect to the scalar product in the space
of wave functionals:

\begin{eqnarray}
&&\left( \Psi _{1},\Psi _{2}\right)  \label{19} \\
&\equiv &\int \prod\limits_{\tau }d^{4}u\left( \tau \right) d^{4}v\left(
\tau \right) \overline{\Psi }_{1}\left[ u\left( \tau \right) ,v\left( \tau
\right) \right] \Psi _{2}\left[ u\left( \tau \right) ,v\left( \tau \right) %
\right] .  \notag
\end{eqnarray}

The new realization of the basic dynamical variables in a space of wave
functionals permits us also to determine an action operator in this space:

\begin{eqnarray}
\widehat{I} &\equiv &\int\limits_{0}^{1}\left[ \frac{\widetilde{\hslash }}{i}%
\overset{\cdot }{u}^{\mu }\frac{\delta }{\delta u^{\mu }}+\frac{\widetilde{%
\hslash }}{i}\overset{\cdot }{v}^{\mu }\frac{\delta }{\delta v^{\mu }}\right.
\notag \\
&&\left. -\left( N_{1}\widehat{H}_{1}+N_{2}\widehat{H}_{2}+\lambda \widehat{D%
}\right) \right] d\tau ,  \label{20}
\end{eqnarray}%
where operators of the constraints are now obtained by substitution of Eqs. (%
\ref{15}) and (\ref{16}) into Eq. (\ref{11}). The operator $\widehat{I}$ is
also formally Hermitian with respect to the scalar product (\ref{19}).

Let us begin formulation of the quantum action principle. It is useful to
re-formulate the eigenvalue problem (\ref{1}), introducing for any wave
functional $\Psi $\ a functional:

\begin{equation}
\Lambda \left[ u,v\right] \equiv \frac{\widehat{I}\Psi \left[ u,v\right] }{%
\Psi \left[ u,v\right] }.  \label{21}
\end{equation}%
For the wave functional the exponential representation:

\begin{equation}
\Psi \left[ u,v\right] \equiv \exp \left( \frac{i}{\widetilde{\hslash }}S%
\left[ u,v\right] +R\left[ u,v\right] \right)  \label{22}
\end{equation}%
will be useful, in particular, to obtain the classical limit $\widetilde{%
\hslash }\rightarrow \infty $. We suppose that the functionals $S \left[ u,v%
\right] ,R\left[ u,v\right] $ are real and analytical, i.e., they are
represented by functional series, for example,

\begin{eqnarray}
&&S\left[ u,v\right]  \notag \\
&=&\int\limits_{0}^{1}\left[ S_{1\mu }\left( \tau \right) u^{\mu }\left(
\tau \right) N_{1}\left( \tau \right) +T_{1\mu }\left( \tau \right) v^{\mu
}\left( \tau \right) N_{2}\left( \tau \right) \right] d\tau  \notag \\
&&+\frac{1}{2}\int\limits_{0}^{1}d\tau \int\limits_{0}^{1}d\tau ^{\prime }%
\left[ S_{2\mu \nu }\left( \tau ,\tau ^{\prime }\right) u^{\mu }\left( \tau
\right) u^{\nu }\left( \tau ^{\prime }\right) N_{1}\left( \tau \right)
N_{1}\left( \tau ^{\prime }\right) \right.  \notag \\
&&+T_{2\mu \nu }\left( \tau ,\tau ^{\prime }\right) v^{\mu }\left( \tau
\right) v^{\nu }\left( \tau ^{\prime }\right) N_{2}\left( \tau \right)
N_{2}\left( \tau ^{\prime }\right)  \notag \\
&&\left. +V_{2\mu \nu }\left( \tau ,\tau ^{\prime }\right) u^{\mu }\left(
\tau \right) v^{\nu }\left( \tau ^{\prime }\right) N_{1}\left( \tau \right)
N_{2}\left( \tau ^{\prime }\right) \right] +  \notag \\
&&...,  \label{23}
\end{eqnarray}%
plus a similar representation for $R\left[ u,v\right] $. The coefficients $%
S_{1\mu },T_{1\mu },S_{2\mu \nu },T_{2\mu \nu },V_{2\mu \nu },...$ are real
functions of the parameter $\tau $, there transformation properties ensure
the invariance of Eq. (\ref{23}) with respect to Lorentz rotations in the
Minkowsky space and reparametrizations of world lines of particles.

Let us formulate conditions of equality of the functional (\ref{21}) to an
eigenvalue of the action operator. These conditions must be applied to
coefficients of the series (\ref{23}). First of all, an eigenvalue must be
independent on internal points $\left( u\left( \tau \right) ,v\left( \tau
\right) \right) $ of a trajectory. It can only depend on the end points $%
\left( u_{0},v_{0}\right) $ and $\left( u_{1},v_{1}\right)$ which are fixed.
Then, we assume that eigenvalues of the action operator are real. This gives
additional conditions on the coefficients of the series. As we shall see,
all these conditions give a set of first order differential equations in the
parameter $\tau \in \left[ 0,1\right] $ for the coefficients $S_{1\mu
},T_{1\mu },S_{2\mu \nu },T_{2\mu \nu },V_{2\mu \nu },...$. This set is an
analog of the Schr\"{o}dinger equation in ordinary quantum mechanics. In the
framework of Cauchy problem, a solution of this set depends on the initial
data at the moment $\tau =0$. This solution also depends functionally on
lapse and schift functions. Substituting this solution into Eq. (\ref{21}),
we obtain an eigenvalue of the action operator as a function (functional) of
enumerated parameters. Therefore, the eigenvalue problem (\ref{1}) is
solved. All enumerated parameters are free and have to be fixed by an
additional condition. The demand of stationarity of the eigenvalue function
(functional) with respect to all free parameters gives a missing condition.
Precisely this condition forms the content of the quantum action principle.
We want to note, that not all of free parameters will be fixed by this
stationarity condition. In this case the stationary eigenvalue $\Lambda _{0}$
of the action operator is degenerate \cite{GL1}. This remark also refers to
lapse and schift functions.

The stationary eigenvalue $\Lambda _{0}$ depends on the end points $\left(
u_{0},v_{0}\right) $ and $\left( u_{1},v_{1}\right)$. Therefore, it can be
chosen as a measure of a distance in a configuration space. So, we arrive at
the notion of quantum geometry. The corresponding eigenfunctional $\Psi _{0}%
\left[ u,v\right] $ describes the quantum dynamics of two particles and has
the natural probabilistic interpretation: $\left\vert \Psi _{0}\left[ u,v%
\right] \right\vert ^{2}$ is a probability density of that a system moves
along a trajectory in neighbourhood of a given trajectory $\left( u\left(
\tau \right) ,v\left( \tau \right) \right) $ \cite{GL}. However, the
physical interpretation of the theory can be obtained independently by use
of the stationary eigenvalue $\Lambda _{0}$. Let us take into account a
functional dependence of $\Lambda _{0}$ on "probe" fields $A^{1,2}$. We
shall consider the quantity $\Lambda _{0}$ as a generating function of mean
values of current densities of particles and their mean correlators. In
particular, quantum analogs of (\ref{8}) are:

\begin{equation}
\left\langle j_{1}^{\mu }\left( u\right) \right\rangle \equiv \left. \frac{%
\delta \Lambda _{0}}{\delta A_{\mu }^{1}\left( u\right) }\right\vert
_{A=0},\left\langle j_{2}^{\mu }\left( v\right) \right\rangle \equiv \left.
\frac{\delta \Lambda _{0}}{\delta A_{\mu }^{2}\left( v\right) }\right\vert
_{A=0}.  \label{24}
\end{equation}

In the next section all parameters related the quantum action principle will
be considered in the classical limit $\widetilde{\hslash }\rightarrow 0$.

\section{CLASSICAL LIMIT OF QUANTUM GEOMETRY}

\bigskip Taking into account the exponential representation of the wave
functional (\ref{22}), one can write the functional (\ref{21}) in the
classical limit as follows:

\begin{eqnarray}
\Lambda \left[ u,v\right]  &=&\int\limits_{0}^{1}d\tau \left\{ \overset{%
\cdot }{u}^{\mu }\frac{\delta S}{\delta u^{\mu }}+\overset{\cdot }{v}^{\mu }%
\frac{\delta S}{\delta v^{\mu }}\right.   \label{25} \\
&&+N_{1}\left[ \left( \frac{\delta S}{\delta u}\right) ^{2}-v^{2}\right]
+N_{2}\left[ \left( \frac{\delta S}{\delta v}\right) ^{2}-u^{2}\right]
\notag \\
&&-N_{1}\left( \frac{\delta S}{\delta u^{\mu }}A^{1\mu }+\frac{\partial
A^{1\mu }}{\partial u^{\mu }}-A_{\mu }^{1}A^{1\mu }\right)   \notag \\
&&-N_{2}\left( \frac{\delta S}{\delta v^{\mu }}A^{2\mu }+\frac{\partial
A^{2\mu }}{\partial v^{\mu }}-A_{\mu }^{2}A^{2\mu }\right)   \notag \\
&&\left. -\lambda \left[ \left( \frac{\delta S}{\delta u^{\mu }}-A_{\mu
}^{1}\right) u^{\mu }-\left( \frac{\delta S}{\delta v^{\mu }}-A_{\mu
}^{2}\right) v^{\mu }\right] \right\} .  \notag
\end{eqnarray}%
This expression is considerably simplified in the classical limit due to the
absence of the functional $R\left[ u,v\right] $ in the representation (\ref%
{25}). To simplify the subsequent consideration, let us omit the "probe" fields at this stage. In this case only a "harmonic" approximation in the
decomposition (\ref{23}) is sufficient. However, a set of differential
equations for the remaining coefficients is two-parametric and rather
complex for an analytical analysis at this stage. In order to simplify this set let us
assume that remaining second-order coefficients are local with respect to the parameter $\tau $:

\begin{eqnarray}
S_{2\mu \nu }\left( \tau ,\tau ^{\prime }\right)  &\equiv &S_{2}\left( \tau
\right) \eta _{\mu \nu }\frac{1}{N_{1}\left( \tau \right) }\delta \left(
\tau -\tau ^{\prime }\right) ,  \notag \\
T_{2\mu \nu }\left( \tau ,\tau ^{\prime }\right)  &\equiv &T_{2}\left( \tau
\right) \eta _{\mu \nu }\frac{1}{N_{2}\left( \tau \right) }\delta \left(
\tau -\tau ^{\prime }\right) ,  \label{26}
\end{eqnarray}%
and $V_{2\mu \nu }\left( \tau ,\tau ^{\prime }\right) \equiv 0$. Then

\begin{eqnarray}
&&\Lambda \left[ u,v\right]  \notag \\
&=&\int\limits_{0}^{1}\left[ \left( \overset{\cdot }{u}^{\mu }-\lambda
u^{\mu }\right) \left( S_{1\mu }-S_{2}u_{\mu }\right) \right.  \notag \\
&&+\left( \overset{\cdot }{v}^{\mu }+\lambda v^{\mu }\right) \left( T_{1\mu
}+T_{2}v_{\mu }\right)  \notag \\
&&+N_{1}v^{2}+N_{2}u^{2}-N_{1}\left( S_{1\mu }+S_{2}u_{\mu }\right) \left(
S_{1}^{\mu }+S_{2}u^{\mu }\right)  \notag \\
&&\left. -N_{2}\left( T_{1\mu }+T_{2}v_{\mu }\right) \left( T_{1}^{\mu
}+T_{2}v^{\mu }\right) \right] d\tau .  \label{27}
\end{eqnarray}%
Integrating by parts the first two terms under the integral, we eliminate
velocities $\overset{\cdot }{u},\overset{\cdot }{v}$. In the remaining
integral, we assume that coefficients in front of the first and second
orders in $u^{\mu }\left( \tau \right)$ and $v^{\mu }\left( \tau \right) $
are equal to zero. The latter condition gives us a set of differential
equations:

\begin{eqnarray}
\overset{\cdot }{S}_{1\mu }-\lambda S_{1\mu }+2N_{1}S_{2}S_{1\mu } &=&0,
\notag \\
\overset{\cdot }{S}_{2}-\lambda S_{2}+2N_{1}S_{2}^{2}-2N_{2} &=&0,  \notag \\
\overset{\cdot }{T}_{1\mu }-\lambda T_{1\mu }+2N_{2}T_{2}T_{1\mu } &=&0,
\notag \\
\overset{\cdot }{T}_{2}-\lambda T_{2}+2N_{2}T_{2}^{2}-2N_{1} &=&0 ,
\label{28}
\end{eqnarray}%
and the remaining nonzero part of Eq. (\ref{27}) becomes equal to an
eigenvalue of the action operator:

\begin{eqnarray}
\Lambda &=&\left. u^{\mu }\left( S_{1\mu }+\frac{1}{2}S_{2}u_{\mu }\right)
\right\vert _{0}^{1}  \notag \\
&&+\left. v^{\mu }\left( T_{1\mu }+\frac{1}{2}T_{2}v_{\mu }\right)
\right\vert _{0}^{1}  \notag \\
&&-\int\limits_{0}^{1}\left( S_{1}^{2}N_{1}+T_{1}^{2}N_{2}\right) d\tau .
\label{29}
\end{eqnarray}

Formulation of the quantum action principle is now extremely transparent.
The solution of the Cauchy problem for the system (\ref{28}), being a
function of initial parameters $S_{1\mu }^{\left( 0\right) },S_{2}^{\left(
0\right) },T_{1\mu }^{\left( 0\right) },T_{2}^{\left( 0\right) }$, and a
functional of the lapse $N_{1,2}\left( \tau \right) $ and schift $\lambda
\left( \tau \right) $ functions, must be substituted in Eq. (\ref{29}). As
the result, one obtains an eigenvalue of the action operator as a function
(functional) of all enumerated parameters. At the final step we have to take
the extremum of that function (functional) with respect to all enumerated
free parameters. The stationary value $\Lambda _{0}$\ of that eigenvalue
function (functional) is a function of only the end points $\left(
u_{0},v_{0}\right) $ and $\left( u_{1},v_{1}\right) $ of a trajectory. It is
this function which is associated with a measure of distance in a quantum
geometry of configuration space. Let us remember, that here we consider only
the classical limit $\widetilde{\hslash }\rightarrow 0$. The following
question naturally arises: is the classical limit of the quantum geometry $%
\Lambda _{0}$ equal to the classical distance $I_{01}$ defined in the first
section? Within the local approximation (\ref{26}) the answer is the following. The classical
limit of the quantum geometry corresponds to an approximation of the classical
Finsler geometry (\ref{5}) for a special choice: $\lambda =0$.
In general case we have no answer on this question. This will be a subject
of a next work. However, an evidence for this equality is that in the
case of one particle the equality of two measures of distance takes place
\cite{GL1}.

Let us return to the problem of determination of observables. "Switching on"
"probe" fields modifies the previous consideration as follows. Let "probe"
fields are real-analytical functions, i.e., they are represented by series:

\begin{eqnarray}
A_{\mu }^{1}\left( u\right) &=&\alpha _{0\mu }^{1}+\alpha _{1\mu \nu
}^{1}u^{\nu }+\frac{1}{2}\alpha _{2\mu \nu \gamma }^{1}u^{\nu }u^{\gamma
}+...,  \notag \\
A_{\mu }^{2}\left( v\right) &=&\alpha _{0\mu }^{2}+\alpha _{1\mu \nu
}^{2}v^{\nu }+\frac{1}{2}\alpha _{2\mu \nu \gamma }^{2}v^{\nu }v^{\gamma
}+...  \label{30}
\end{eqnarray}%
In this case, the "harmonic" approximation for functionals $S\left[ u,v%
\right] ,R\left[ u,v\right] $ is not enough. The set of four differential
equations (\ref{28}) is replaced by an infinite set of differential
equations with coefficients which depend on the infinite set of field
parameters $\alpha _{0\mu }^{1,2},\alpha _{1\mu \nu }^{1,2},\alpha _{2\mu
\nu \gamma }^{1,2},...$. Let us assume that in this case the Cauchy problem
also has a smooth solution. Let the corresponding eigenvalue functional has
a unique stationary value $\Lambda _{0}$ with respect to all free
parameters. This quantity plays the role of a generating function of
physical observables. In particular, for mean value of current density of
the first particle one obtains the infinite set of equations:

\begin{eqnarray}
\left. \frac{\delta \Lambda _{0}}{\delta \alpha _{0\mu }^{1}}\right\vert
_{\alpha =0} &=&\int d^{4}u\left\langle j_{1}^{\mu }\left( u\right)
\right\rangle ,  \notag \\
\left. \frac{\delta \Lambda _{0}}{\delta \alpha _{1\mu \nu }^{1}}\right\vert
_{\alpha =0} &=&\int d^{4}u\left\langle j_{1}^{\mu }\left( u\right)
\right\rangle u^{\nu },  \notag \\
\left. \frac{\delta \Lambda _{0}}{\delta \alpha _{0\mu \nu \gamma }^{1}}%
\right\vert _{\alpha =0} &=&\int d^{4}u\left\langle j_{1}^{\mu }\left(
u\right) \right\rangle u^{\nu }u^{\gamma },....  \label{31}
\end{eqnarray}%
where all integrals are over $M_{4}$.

\section{\textbf{CONCLUSIONS}}

In conclusion, the quantum action principle gives us a possibility to derive
a correct quantum version of a covariant dynamical theory with a proper
probabilistic interpretation. It also gives us an alternative tool for
determination of observables. In the presence of "probe" fields, it is the
stationary eigenvalue of the quantum action that plays the role of a
generation function of currents and their correlators.

We are thanks V. A. Franke and A. V. Goltsev for useful discussions.




\end{document}